\documentclass[%
reprint,
amsmath,amssymb,
aps,
]{revtex4-2}

\usepackage{graphicx}
\usepackage{dcolumn}
\usepackage{bm}

\usepackage{braket}
\usepackage{graphicx}
\usepackage{amsmath}
\usepackage{color}
\begin{document}

\title{Rydberg atomic spectrum analyzer with microwave-dressed-state-locking and multimode Floquet theory.}%

\author{Sheng-Xian Xiao}
\affiliation{Chongqing Key Laboratory for Strongly Coupled Physics, Chongqing University, Chongqing, 401331, China}

\affiliation{Center of Modern Physics, Institute for Smart City of Chongqing University in Liyang, Liyang 213300, China}
\author{Tao Wang}
\thanks{corresponding author: tauwaang@cqu.edu.cn}
\affiliation{Chongqing Key Laboratory for Strongly Coupled Physics, Chongqing University, Chongqing, 401331, China}

\affiliation{Center of Modern Physics, Institute for Smart City of Chongqing University in Liyang, Liyang 213300, China}
\begin{abstract}
We propose a Rydberg atomic spectrum analyzer (RASA) utilizing microwave-dressed-state-locking (MWDSL) in conjunction with multimode Floquet theory (MFT). By leveraging a strong local microwave (MW) field resonant with Rydberg states to implement MWDSL, we analyze the second-order effect of MFT induced by the interplay of controllable bias and signal MW fields. This effect facilitates the coupling of locked dressed states, providing a pathway for measuring the signal MW field. We demonstrate that the RASA can simultaneously characterize multiple MW fields across distinct frequencies, with both the frequency and strength of each MW field discernible in the spectral response. This capability renders RASA suitable for measuring unknown-frequency MW fields, thereby expanding the utility of Rydberg atom-based electrometers in complex spectral analysis scenarios.

\end{abstract}
\maketitle
Over the past dozen years, microwave (MW) electric field sensing based on Rydberg atoms has rapidly developed due to their high sensitivity to electric fields \cite{1-Fan_2015,2-Yuan_2023,3-ZHANG20241515}. This technology uses electromagnetically induced transparency (EIT) \cite{4-RevModPhys.77.633} and Autler-Townes (AT) splitting \cite{5-PhysRevA.81.053836} to realize \cite{11-Sedlacek}. The Rydberg atomic sensor can be used both as an electrometer to measure the absolute strength of the MW electric field and as a receiver to detect time-varying signals. The experimental demonstrations of the recovery of time-varying signals with amplitude modulation \cite{AM1-10.1063/1.5028357,AM2-9069423,AM3-Li:22}, frequency modulation \cite{AM2-9069423} or phase modulation \cite{PM-8778739,PM2-PhysRevApplied.19.044079} demonstrate the great potential of Rydberg atomic receivers in communication. Due to its self-calibrating, compact size independent of wavelength, ultra-wide working frequency range \cite{Meyer_2020} and high sensitivity \cite{12-Jing}, this technology has the potential to replace traditional antennas in sensing and communication applications.

However, since Rydberg atoms are only sensitive to near-resonant MW electric fields \cite{12-Jing,13-10.1063/1.4947231,14-Kumar,15-10.1063/5.0069195,16-10.1063/5.0146768}, simultaneously identifying multi-frequency MW electric fields, especially those spanning different bands, remains challenging. Due to the current instantaneous bandwidth limitations of Rydberg atomic receivers \cite{PhysRevApplied.18.034030}, this capability is crucial to improve the data transfer capacity and efficiency in wireless communication and radar systems. While several methods for measuring continuous-frequency MW electric fields have been developed to achieve MHz to THz sensing, such as using far-detuning AC Stark effects \cite{17-PhysRevApplied.15.014053,18-10.1063/5.0086357}, adjacent Rydberg resonance tuning \cite{19-PhysRevA.104.032824,20-PhysRevApplied.19.044049}, two-photon MW transitions \cite{21-10.1063/1.4996234,22-PhysRevApplied.18.054003}, auxiliary MW fields \cite{23-PhysRevA.103.063113,24-PhysRevA.107.043102} and Floquet states \cite{10.1063/5.0227250}, these approaches are typically designed for single-frequency MW electric field or can only measure MW electric fields with frequency differences within the instantaneous bandwidth. Some researches have attempted to use double the number of
lasers \cite{9054945}, multiple Rydberg states \cite{PhysRevApplied.19.014025} or spatial division multiplexing \cite{ZHANG2024100089} to receive MW electric fields from different bands simultaneously. Whereas, the number of local oscillator (LO) fields or lasers required by these methods needs to increase with the number of MW signal fields, boosting system complexity and cost, which is detrimental to commercialization for this practical technology.

In this letter, we design a Rydberg atomic spectrum analyzer (RASA) for simultaneously measuring and identifying multi-frequency microwave signals with only one LO MW field required. Previous approaches solely depended on the energy splitting of MW-dressed states for electric field measurement \cite{11-Sedlacek,12-Jing}. In contrast, our method utilizes the coupling between these states, which was not achieved in prior researches. This coupling interaction stems from the quantum mixing effect between the tunable bias MW field and the signal MW field, interpretable via multimode Floquet theory (MFT) \cite{25-PhysRevX.12.021061}. As the MW-dressed states are locked by the strong LO field, the system only responds when the mixing effect resonates with the MW-dressed states. By scanning the frequency of the  bias field, our method enables the measurement and identification of MW signals with distinct frequencies. 

\begin{figure}
	\includegraphics[width=1\linewidth]{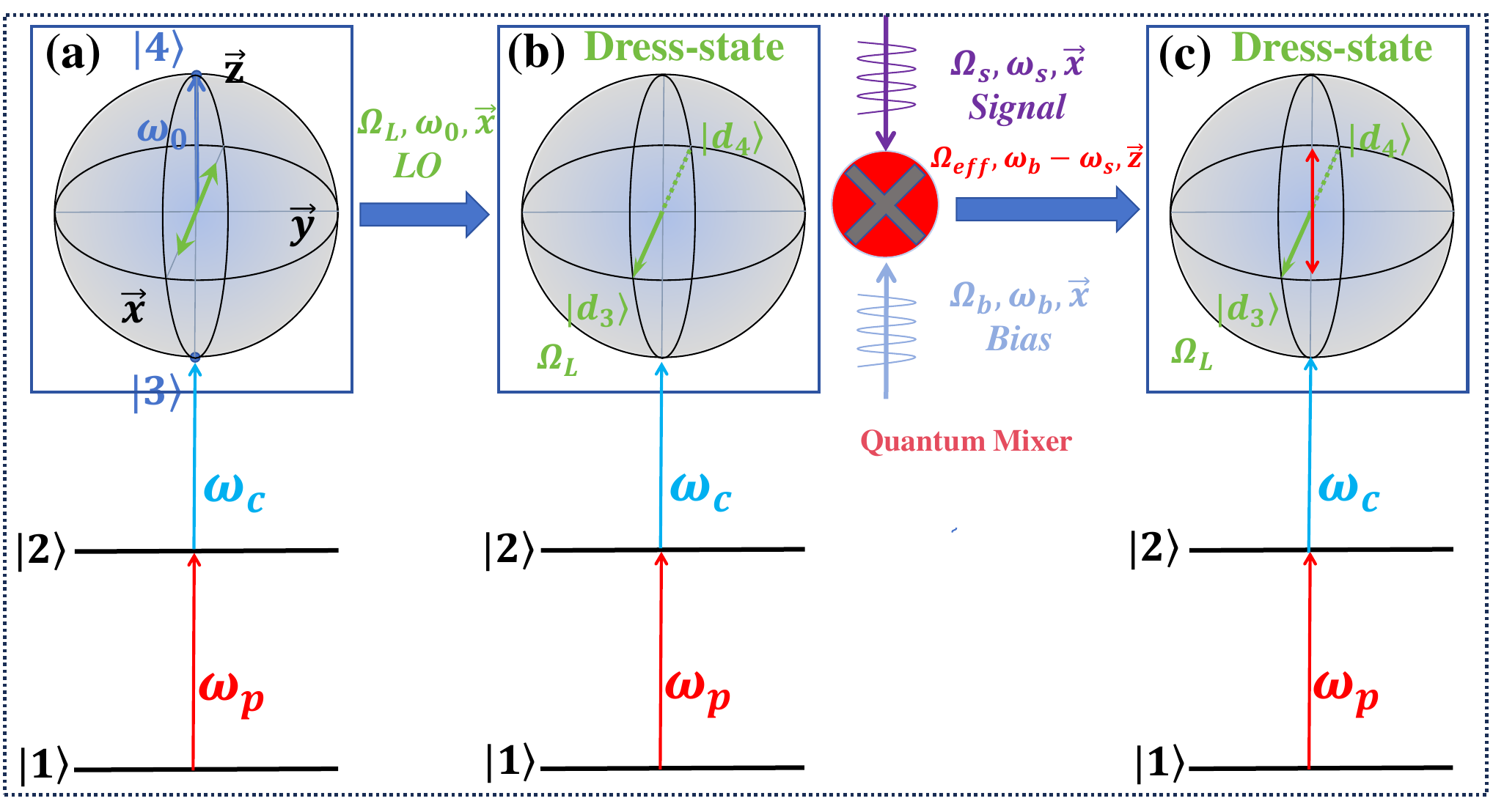}
        \includegraphics[width=1\linewidth]{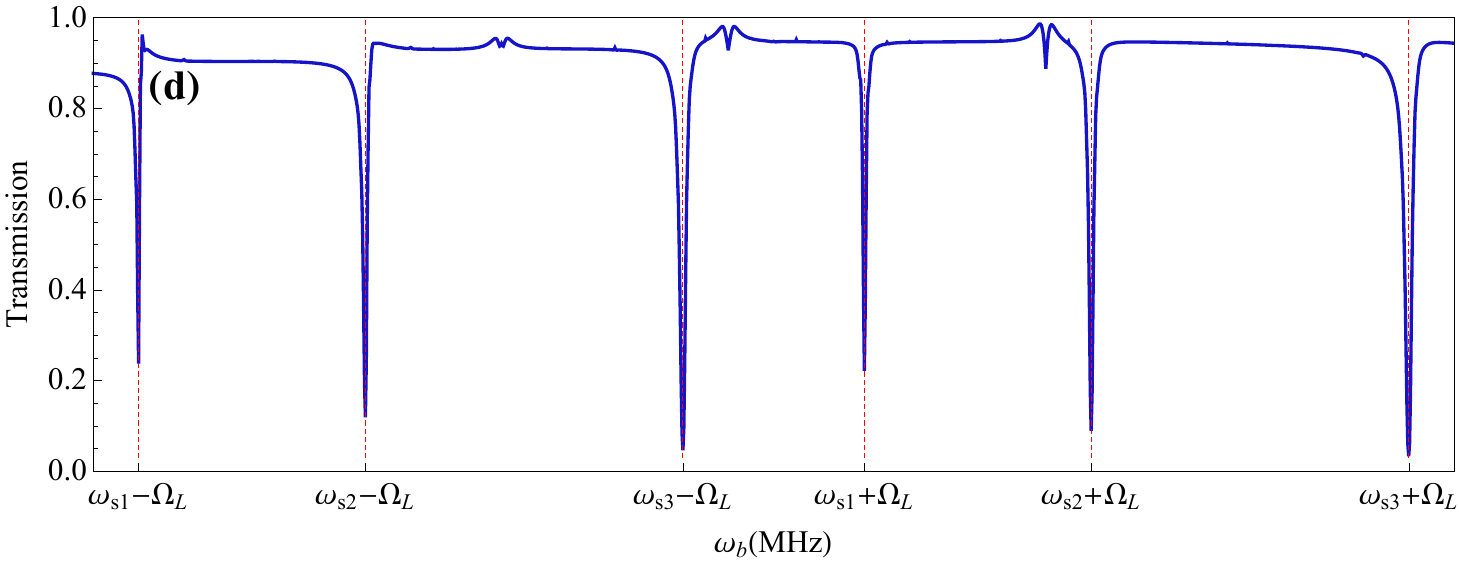}
	\caption{(a)-(c) Schematic diagram of the principle of RASA. (a) is a four-level system, in which two Rydberg states are represented as the eigenstates in the z-direction of the Bloch sphere. The LO field in the $x$ direction induces the formation of the MW-dressed-states in (b). The signal field and bias field both in $x$ direction form an effective field in the $z$ direction through a quantum mixer, coupling the MW-dressed-states to form a new four-level system in (c). (d) The transmission spectrum when three signals exist simultaneously, with parameters set as  $\Omega_L/(2\pi)=80$ MHz, $\Delta_c/(2\pi)=40$ MHz, $\Delta_{s_1}/(2\pi)=1000$ MHz, $\Delta_{s_2}/(2\pi)=1050$ MHz, $\Delta_{s_1}/(2\pi)=1120$ MHz, $\Omega_{s_1}/(2\pi)=7.5$ MHz, $\Omega_{s_2}/(2\pi)=10$ MHz, $\Omega_{s_1}/(2\pi)=12.5$ MHz. The scanning method is fixing $\Omega_b/\Delta_b=$0.1.\label{fig1}}
\end{figure}

The RASA is a typical Rydberg atom-based electrometer with a four-level system consisting of ground state $\left| 1\right\rangle$, excited state $\left| 2\right\rangle $, and Rydberg states $\left| 3\right\rangle $, $\left| 4\right\rangle $, depicted in Figs. \ref{fig1}(a)-(c). The weak probe laser, with Rabi frequency $\Omega_p$, is resonant with transition $\left| 1\right\rangle \longleftrightarrow  \left| 2\right\rangle$. The control laser, with Rabi frequency $\Omega_c$ and detuning $\Delta_c$, is coupled to the transition $\left| 2\right\rangle \longleftrightarrow  \left| 3\right\rangle$. A strong LO MW field, a bias MW field, and weak signal MW fields are simultaneously coupled to the transition $\left| 3\right\rangle \longleftrightarrow  \left| 4\right\rangle$. Their Rabi frequencies are $\Omega_L$, $\Omega_b$ and $\Omega_{s_i}$ (where $i=1,2,3...$) respectively, and their corresponding frequencies are $\omega_0$, $\omega_b$ and $\omega_{s_i}$. To offer a more intuitive understanding, we represent this process using the Bloch sphere, commonly applied in two-level system. Here, Rydberg states $\left| 3\right\rangle $ and $\left| 4\right\rangle $ serve as eigenstates of the $z$ direction, exhibiting a frequency difference of $\omega_0$. The LO field achieves resonance, whereas the bias and signal fields are off-resonant. All these fields are oriented along along the $x$ direction in the Bloch sphere. The LO field generates two MW-dressed states, which are also eigenstates of the $x$ direction, with a frequency difference of $\Omega_L$. Denoted as $\left| d_3\right\rangle $ and $\left| d_4\right\rangle$, these MW-dressed states can form two three-level EITs: $\left| 1\right\rangle \longleftrightarrow  \left| 2\right\rangle \longleftrightarrow  \left| d_3\right\rangle$ and $\left| 1\right\rangle \longleftrightarrow  \left| 2\right\rangle \longleftrightarrow  \left| d_4\right\rangle$. By setting $\Delta_c$ to $\pm\Omega_L/2$, two bright peaks emerge in the transmission spectrum when scanning $\Delta_c$, a phenomenon known as EIT-AT splitting. The bias and signal field, through the quantum mixer, create an effective field in the $z$ direction with a frequency $\omega_b-\omega_{{s_i}}$, which can be coupled to the transition $\left|d_3\right\rangle \longleftrightarrow  \left| d_4\right\rangle$. By setting $\omega_b$ to satisfy $|\omega_b-\omega_{{s_i}}|=\Omega_L$ and locking the detuning of the control laser at $\Omega_L/2$ ($-\Omega_L/2$), a four-level configuration $\left| 1\right\rangle \longleftrightarrow  \left| 2\right\rangle \longleftrightarrow  \left| d_3\right\rangle \longleftrightarrow  \left| d_4\right\rangle$ ($\left| 1\right\rangle \longleftrightarrow  \left| 2\right\rangle \longleftrightarrow  \left| d_4\right\rangle \longleftrightarrow  \left| d_3\right\rangle$), similar to the Rydberg electrometer \cite{11-Sedlacek}, is achieved, enabling the measurement of the signal fields. In fact, in the picture of the MW-dressed states, the control laser simultaneously couples $\left| 2\right\rangle $ with $\left| d_3\right\rangle $ and $\left| d_4\right\rangle $, with a coupling strength of $\Omega_c/\sqrt{2}$ for both. This implies the concurrent existence of the two four-level configurations. However, when $\Delta_c=\pm\Omega_L/2$ is selected, one configuration has a detuning of $\Omega_L$, while the other has zero detuning. Thus, for condition $\Omega_L\gg\Omega_c/\sqrt{2}$, typically satisfied for a strong LO field, we only need to consider the zero-detuning configuration.  

Multiple MW signals can also generate fields in the z-direction via a quantum mixer. However, two factors prevent the mixing between signals from affecting the measurement. First, mixing between signals may not satisfy the resonance of the dressed states. Second, due to the weak signals, their mixing effect is much smaller than that with the bias field. Consequently, by scanning the bias frequency in an appropriate way and measuring the transmission spectrum of the probe laser, frequencies and strengths of multiple signals can be simultaneously determined, as shown in Fig. \ref{fig1}(d). It should be noticed that in the absence of the signal field, the transmission of the probe laser corresponds to one of the transparent peaks of the LO field. Thus, the emergence of a signal field manifests as absorption in the transmission spectrum. As can be seen Fig. \ref{fig1}(d), each signal's response in the transmission spectrum appears as two absorption peaks. The difference between the bias field frequencies corresponding to the two absorption peaks is $2\Omega_L$, and their average exactly matches the frequency of the signal. Additionally, the stronger the signal, the higher the absorption peaks. Thus, RASA can simultaneously measure multiple signals' frequencies and strengths and can also be used to search for unknown-frequency signals.

When multiple signals with different frequencies are simultaneously present, the original Hamiltonian of the RASA with rotating-wave approximation takes the form: $H(t)=-\hbar\Delta_c\left(\left| 3\right\rangle \left\langle 3\right|+\left| 4\right\rangle \left\langle 4\right|\right)-\frac{\hbar}{2}\left(\Omega_p\left| 1\right\rangle \left\langle 2\right|+\Omega_c\left| 2\right\rangle \left\langle 3\right|+h.c.\right)+H_B(t)$, where $H_B(t)$ represents the Hamiltonian in the Bloch sphere and  is given by: $H_B(t)=-\frac{\hbar}{2}\left(\Omega_L+\Omega_{b}e^{i\Delta_{b}t}+\sum_i\Omega_{s_{i}}e^{i\Delta_{s_{i}}t}\right)\sigma_-+h.c.$. Here, $\hbar$ is reduced Planck's constant. $\Delta_b=\omega_b-\omega_0$ and $\Delta_{s_{i}}=\omega_{s_{i}}-\omega_0$ represent the detunings of the bias and signal fields, respectively. $\sigma_i$ (i=$x,y,z$) are the Pauli matrices and $\sigma_-=\left| 3\right\rangle \left\langle 4\right|$. By applying MFT (see Appendix.\ref{ap1}), the effective field of the $z$ direction in the Bloch sphere is calculated as  
$-\hbar\Omega_{s_{i}}^{eff}\cos\left[(w_b-\omega_{s_{i}})t\right]\sigma_z$, where the effective Rabi frequency satisfies $\Omega_{s_{i}}^{eff}=\frac{\Omega_b\Omega_{s_{i}}}{4}\left(\frac{1}{\Delta_b}+\frac{1}{\Delta_{s_i}}\right)$. In addition to this, the bias field also induces a Stark effect, represented as $-\hbar\delta\sigma_z$, where $\delta=\Omega_b^2/(4\Delta_b)$. As a result, the effective Hamiltonian in the Bloch sphere is given by: $H^{eff}_B(t)=-\frac{\hbar}{2}\Omega_L\sigma_x-\hbar\left[\sum_i\Omega_{s_{i}}^{eff}\cos(w_b-\omega_{s_{i}})t+\delta\right]\sigma_z$. In our calculation, we have assumed that $\Delta_b$ and $\Delta_{s_i}$ share the same sign. Moreover, the Stark effect induced by the signal fields has been neglected due to their relatively weak strengths. Given that the LO field is extremely strong ($\Omega_L\gg\delta$), the Stark effect can be disregarded under the resonant condition ($|\omega_b-\omega_{s_i}|=\Omega_L$). Therefore, the total effective Hamiltonian of the system can be expressed as: $H^{eff}(t)=-\hbar\Delta_c\left(\left| 3\right\rangle \left\langle 3\right|+\left| 4\right\rangle \left\langle 4\right|\right)
        -\frac{\hbar}{2}\left(\Omega_p\left| 1\right\rangle \left\langle 2\right|+\Omega_c\left| 2\right\rangle \left\langle 3\right|+h.c.\right)-\frac{\hbar}{2}\Omega_L\sigma_x-\hbar\left[\sum_i\Omega_{s_{i}}^{eff}\cos(w_b-\omega_{s_{i}})t\right]\sigma_z$. 

To obtain the transmission spectrum of the signals, we simulate the system dynamic numerically using the master equation $\dot{\rho}=-\frac{i}{\hbar}[H,\rho]+\mathcal{L}(\rho)$, where $\mathcal{L}(\rho)$ is the Lindblad operator that accounts for the decay processes in the atom \cite{31-10.1063/1.4984201}. The transmission of the probe laser is determined by the density matrix component $\rho_{21}$ via: $T=\exp{\left[-\mathcal{K} \mathrm{Im}(\rho_{21})\right]}$, where $\mathcal{K}$ depends on parameters such as the probe laser intensity and wavelength, the atomic gas density, the length of the vapor cell and the transition dipole moment between states $\left| 1\right\rangle$ and $\left| 2\right\rangle$. In this paper, we set $\mathcal{K}=50$, $\Omega_p/(2\pi)=2$ MHz, $\Omega_c/(2\pi)=2$ MHz, and the state decay parameters $\Gamma_2/(2\pi)=5$ MHz, $\Gamma_3=2$ kHz, $\Gamma_4=1$ kHz.

\begin{figure}
	\centering
	\includegraphics[width=1\linewidth]{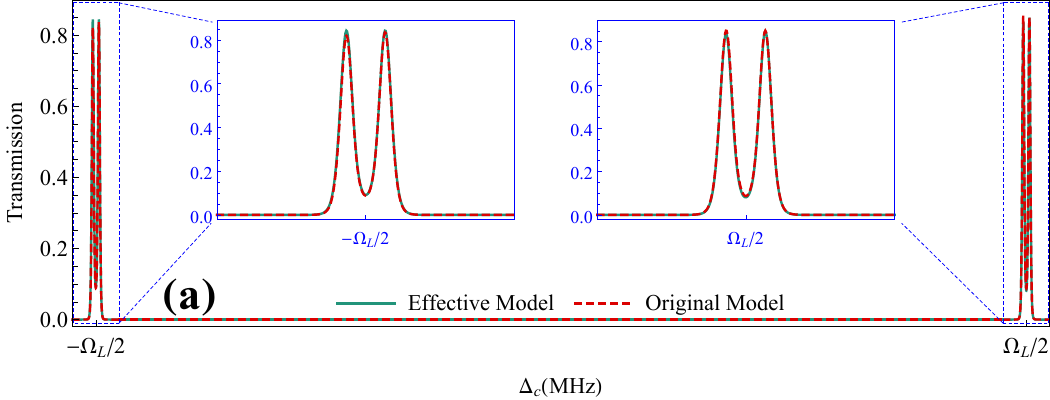}
 	\includegraphics[width=1\linewidth]{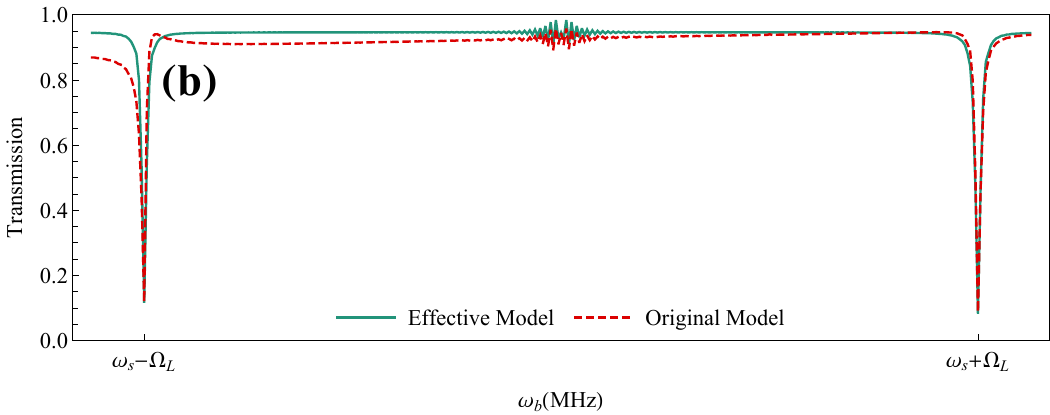}
	\caption{Transmission spectra calculated using the original and effective Hamiltonians. (a) Transmission spectrum of scanning $\Delta_c$ with $\Delta_b-\Delta_s=\Omega_L$. (b) Transmission spectrum of scanning $\Delta_b$ with $\Delta_c=\Omega_L/2$. The parameters are set as $\Delta_s/(2\pi)=1000$ MHz, $\Omega_L/(2\pi)=80$ MHz. \label{fig2}}
\end{figure}

Here, we perform a proof of principle without considering the Doppler effect. As shown in Fig. \ref{fig2}, we present the transmission spectra calculated using the original and effective Hamiltonians, with a single signal field considered. In Fig. \ref{fig2}(a), we set $\omega_s-\omega_b=\Omega_L$, achieving resonance of the effective field with the dressed-state transition. As can be seen, the effective model's results align perfectly with those of the original model. Both AT-splitting peaks of the LO field undergo further splitting, forming the AT-splitting of the effective field, with a magnitude of exactly $\Omega_s^{eff}$. This substantiates the validity of employing the Bloch sphere and dressed states to elucidate the RASA mechanism. Moreover, it indicates that the factor $\frac{\Omega_b}{4}\left(\frac{1}{\Delta_b}+\frac{1}{\Delta_{s_i}}\right)$ quantifies the reduced sensitivity of RASA to electric fields compared to the typical Rydberg atom-based electrometer. 

\begin{figure}
	\centering
        \includegraphics[width=1\linewidth]{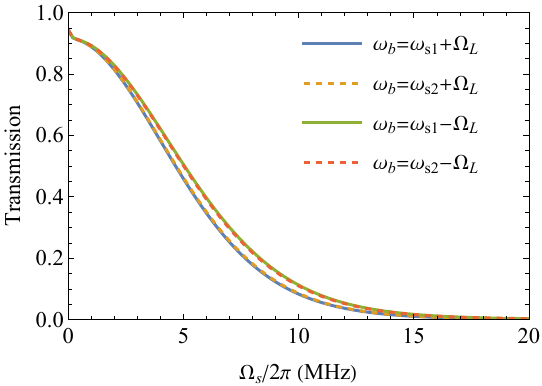}
	\caption{The transmission varies with the signal strength under resonant conditions. The parameters are set as $\Delta_{s_1}/(2\pi)=1000$ MHz, $\Delta_{s_2}/(2\pi)=1200$ MHz, $\Omega_L/(2\pi)=80$ MHz, $\Delta_c=\Omega_L/2$ and $\Omega_b/\Delta_b=0.1$.\label{fig3}}
\end{figure}

To search for the signal, we fix $\Delta_c$ at $\Omega_L/2$, corresponding to a four-level configuration of $\left| 1\right\rangle \longleftrightarrow  \left| 2\right\rangle \longleftrightarrow  \left| d_3\right\rangle \longleftrightarrow  \left| d_4\right\rangle$, and scan the bias field with $\Omega_b/\Delta_b=0.1$ to obtain the transmission spectrum. As shown in Fig. \ref{fig2}(b), both the original and effective Hamiltonians exhibit absorption peaks at $\omega_b=\omega_s\pm\Omega_L$ in their transmission spectra, which are identical. This provides a method for determining the signal frequency by calculating the average of the bias frequencies corresponding to the two absorption peaks.  

Once the signal frequency is determined, its strength can be inferred from the corresponding absorption peak height. As Fig. \ref{fig3} illustrates, stronger signals result in lower transmitted values. While minor differences in the peak heights at $\omega_b=\omega_s\pm\Omega_L$ exist, these differences are likely to be masked by experimental system noise \cite{11-Sedlacek}. Importantly, by fixing $\Omega_b/\Delta_b$, the effective field strength $\Omega_s^{eff}\approx\frac{\Omega_b}{2\Delta_b}\Omega_s$ becomes nearly independent of the signal frequency under the MFT conditions. Fig. \ref{fig3} shows that signals of different frequencies share the same relationship between transmission and strength. Consequently, the strength comparison of signals at different frequencies can be directly reflected in the spectrum, as shown in Fig. \ref{fig1}(d).

\begin{figure*}
    \centering
	\includegraphics[width=1\linewidth]{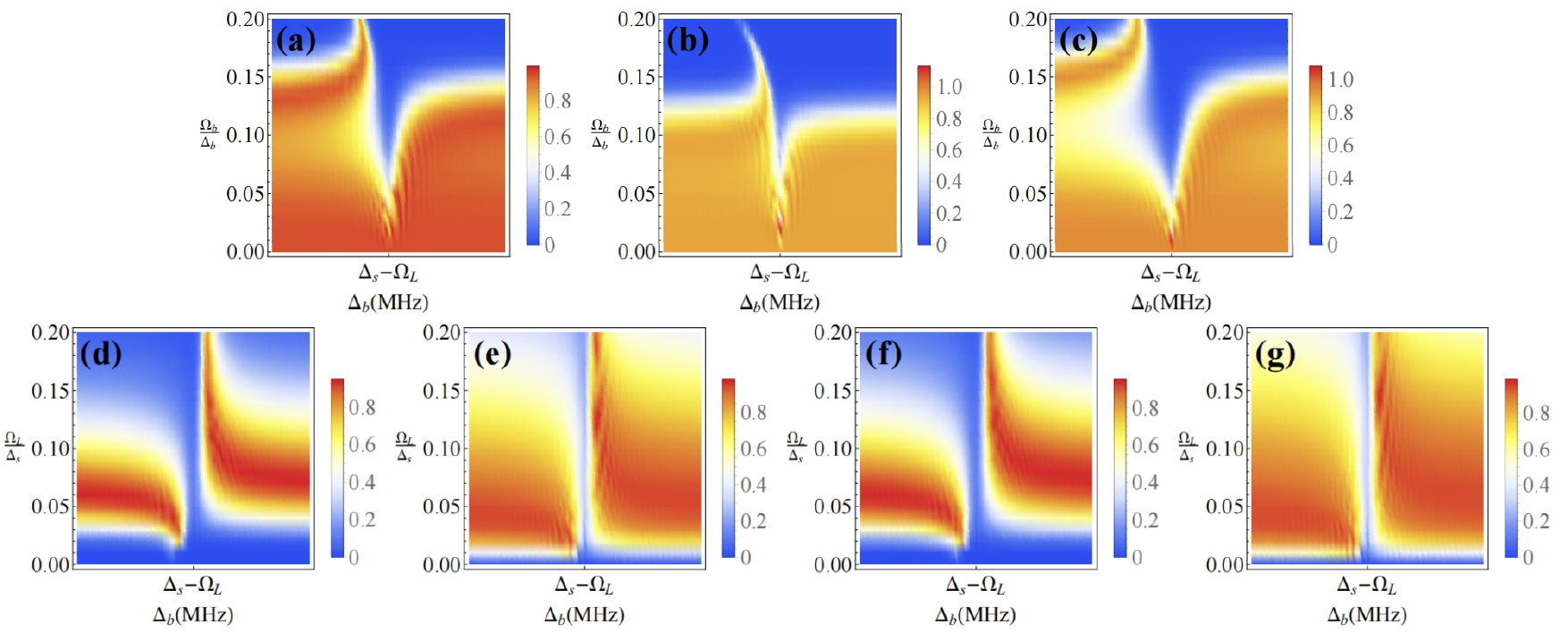}
	\caption{\label{fig4}(a)-(c) Transmission spectra vary with $\Omega_b/\Delta_b$. (a) $\Delta_s/(2\pi)=1000$ MHz, $\Omega_s/(2\pi)=10$ MHz, $\Omega_L/(2\pi)=80$ MHz. (b) $\Delta_s/(2\pi)=1050$ MHz, $\Omega_s/(2\pi)=7.5$ MHz, $\Omega_L/(2\pi)=60$ MHz. (c) $\Delta_s/(2\pi)=1120$ MHz, $\Omega_s/(2\pi)=12.5$ MHz, $\Omega_L/(2\pi)=100$ MHz. (d)-(g) Transmission spectra vary with $\Omega_L/\Delta_s$. The parameters are set as $\Omega_s/(2\pi)=10$ MHz, (a) $\Delta_s/(2\pi)=1000$ MHz, $\Omega_b/\Delta_b=0.1$, (b) $\Delta_s/(2\pi)=1000$ MHz, $\Omega_b/\Delta_b=0.07$, (c) $\Delta_s/(2\pi)=800$ MHz, $\Omega_b/\Delta_b=0.1$, (d) $\Delta_s/(2\pi)=800$ MHz, $\Omega_b/\Delta_b=0.07$}
\end{figure*}

However, as $\omega_b$ gradually moved away from the resonance, discrepancies in absorption between the original and effective Hamiltonians emerge, particularly as $\omega_b$ decreases. Two main factors contribute to this: first, under non-resonant conditions, the long time evolution of the system dynamics can't ignore Stark effect $\delta$, leading to asymmetric spectrum; second, as $\omega_b$ ($\Delta_b$) decreases, MFT begins to show a tendency to fail. Despite these factors, by selecting appropriate parameters, the RASA can still accurately measure the signal's frequency and strength within its frequency measurement range. In our theoretical calculations, we made two approximations: MTF and ignoring the stark effect. The conditions for these approximations are: $\Omega_L,\Omega_b,\Omega_{s_i}\ll\Delta_b,\Delta_{s_i}$ and $\Omega_L\gg\Omega_b^2/(4\Delta_b)$. Since the signal is a weak field, $\Omega_{s_i}\ll\Delta_b,\Delta_{s_i}$ is automatically satisfied. Therefore, these approximate conditions primarily impose limitations on the scanning method of the bias field, the intensity of the LO field, and the frequency range of the detectable signal, which are explored in the following by simulating the transmission spectrum of the original Hamiltonian. 

To explore the relationship between $\Delta_b$ and $\Omega_b$, which determines the scanning method of bias field and the sensitivity of the RASA due to the relation $\frac{\Omega_b}{4}\left(\frac{1}{\Delta_b}+\frac{1}{\Delta_{s_i}}\right)\approx\Omega_b/(2\Delta_b)$, we must consider the following. When multiple signals with unknown frequencies are present, directly scanning $\Delta_b$ while fixing $\Omega_b$ may cause to MFT to fail for small $\Delta_{s_i}$ and result in low sensitivity for lager $\Delta_{s_i}$. Hence, scanning the bias field by fixing the value of $\Omega_b/\Delta_b$ is a better method. This ensures $\Omega_b\ll\Delta_{s_i}(\approx\Delta_b)$ is satisfied near the resonance and maintains high sensitivity for each signal, thereby achieving accurate measurement of multiple signals. Figs. \ref{fig4}(a)-(c) shows how the transmission spectra vary with $\Omega_b/\Delta_b$ when different frequency signals exist independently. As $\Omega_b/\Delta_b$ increases, both absorption peaks at $\Delta_{s}\pm\Omega_L$ become higher, indicating increased sensitivity. However, when $\Omega_b/\Delta_b$ exceeds approximately a critical value, the absorption peaks disappear, meaning signal information can't be obtained from the spectrum. This critical value increases with $\Omega_L$,  suggesting that increasing $\Omega_L$ can enhance sensitivity limits. Yet, $\Omega_L$ can't be infinitely increased and must satisfy $\Delta_s\gg\Omega_L\gg\Omega_b^2/(4\Delta_b)\approx\Delta_s\Omega_b^2/(2\Delta_b)^2$.  As shown in Figs. \ref{fig4}(d)-(g), $\Omega_L/\Delta_s$ has upper and critical values, which are almost the same for different $\Delta_s$. The upper critical value $\alpha_{upper}$ increases and the lower critical value $\alpha_{lower}$ decreases as $\Omega_b/\Delta_b$ decreases. Thus, there's a contradiction in simultaneously increasing the critical value of $\Omega_b/\Delta_b$ and the upper critical value of $\Omega_L$. This reveals a contradiction between expanding the frequency measurement range and improving sensitivity. Based on the analysis, the frequency measurement range of the RASA is given by: $\Omega_L/\alpha_{upper}\le|\Delta_s|\le\Omega_L/\alpha_{lower}$, where $\alpha_{upper}$ and $\alpha_{lower}$ depend on $\Omega_b/\Delta_b$. This relation shows that once the strength of the LO field and the ratio of scanning the bias field are set, the main performance of the RASA is essentially established. Smaller $\Omega_L$ and larger $\Omega_b/\Delta_b$ are better for measuring signals with smaller detuning, while for signals with larger detuning, increasing $\Omega_L$ as much as possible is necessary to maintain high sensitivity. Note that in our theory, the frequency difference $\omega_0$ between Rydberg states is arbitrary. Due to the rich Rydberg energy levels, the RASA's frequency measurement range can be extended to cover terahertz and megahertz electric fields.


In summary, we have proposed a novel type of spectrum analyzer based on the Rydberg atomic system, which is feasible to implement experimentally. We have conducted a proof-of-principle demonstration of the spectrum analyzer and explored the parameter conditions for the LO field used to lock the dressed-states and the bias field used for scanning. Our research can be extended to multi-frequency MW communication, in which the number of MWs exists in pairs. As long as the frequency difference is $\Omega_L$, each pair of MWs can originate from different frequency bands. Furthermore, similar to applying a multi-pulse sequence in a two-level system, changing the way the LO field is applied to manipulate the dressed-states might lead to a new method for measuring the MW electric field \cite{25-PhysRevX.12.021061,RevModPhys.89.035002}.
\begin{acknowledgments}
This work is supported by National Science Foundation of China under Grant No. 12274045, No. 12274046 and No.12347101.
\end{acknowledgments}

\appendix
\section{quantum frequency mixing base on MFT \label{ap1}}
We consider a time-dependent Hamiltonian that can be described by two frequency modes $(\omega_a,\omega_b)$: 
\begin{eqnarray}
	H(t)=\sum_{m,n}H_{m,n}e^{im\omega_at}e^{im\omega_bt},\label{eq3}
\end{eqnarray}
where $\omega_a$ and $\omega_b$ are much larger than other energy parameters. According to the theory of the quantum frequency mixing, the time-dependent effective Hamiltonian with low-frequency can be obtained as \cite{25-PhysRevX.12.021061}:
\begin{eqnarray}
	H^{eff}(t)=\sum_{l,k}\left( H_{l,k}+H^{(2)}_{l,k}+\dots\right)e^{i(l\omega_a+k\omega_b)t},\label{eq4}
\end{eqnarray}
where summation indices $(l,k)$ satisfy the condition that $l\omega_a+k\omega_b\ll\omega_a,\omega_b$ is a low frequency. The second-order term $H^{(2)}_{l,k}$ is given by:
\begin{eqnarray}
	-\frac{1}{2}\sum_{(p,q)\neq (l,k)}\frac{\left[ H_{l-p,k-q},H_{p,q}\right] }{p\omega_a+q\omega_b},\label{eq5}
\end{eqnarray}
where the summation excludes the low-frequency case $(p,q)=(l,k)$. Non-commutation in second-order terms enables quantum frequency mixing: the non-commutation between high-frequency terms is crucial to quantum mixing theory. This theory can be generalized to the case of multiple frequency modes.

\bibliographystyle{apsrev4-2}
\bibliography{ref}

\end{document}